\documentclass[prl,twocolumn]{revtex4-1}

\usepackage{graphicx}
\usepackage{multirow}
\usepackage{dcolumn}
\usepackage{bm}
\usepackage[normalem]{ulem}
\usepackage{amsmath}
\usepackage{amsfonts}
\usepackage{amssymb}
\usepackage{color}
\usepackage{colordvi}
\usepackage{dcolumn}
\usepackage{bm}

\usepackage[colorlinks=true, breaklinks=true, linkcolor=blue,
urlcolor=blue, citecolor=blue]{hyperref}
\renewcommand\Re{\operatorname{Re}}
\renewcommand\Im{\operatorname{Im}}
\DeclareMathOperator{\Tr}{Tr}

\begin{document}

\author{Yury Sherkunov}
\email{sherkunov@gmail.com} 
\affiliation{Department of Physics, Loughborough University, Loughborough LE11 3TU, United Kingdom}

\title{ Resonance oscillations of non-reciprocal long-range van der Waals forces between  atoms in electromagnetic fields}

\date{\today }

\begin{abstract}
We study theoretically the van der Waals interaction between two atoms  out of equilibrium with isotropic electromagnetic field. We demonstrate that at large interatomic separations, the van der Waals forces are resonant, spatially oscillating and non-reciprocal due to resonance absorption and emission of virtual photons.   We suggest that the van der Waals forces can be controlled and manipulated by tuning the spectrum of artificially created random light.
\end{abstract}

\maketitle

\section{Introduction}

The long-range dispersion interaction between atoms arising from quantum or thermal fluctuations  of electromagnetic (EM) field  and atomic charges has been well understood  for equilibrium systems since pioneering  works by Casimir and Polder \cite{CasimirPolder} and Lifshitz with collaborators  \cite{Lifshitz, Lifshitz61}. The situation is different for non-equilibrium systems, where, \textit{e.g.}, the long-distance   dispersion interaction between an excited and ground-state atoms  has been a subject of intense theoretical debate for nearly fifty years \cite{McLone65, Power95, Sherkunov09,Safari15, Berman15, Donaireetal15, Milonni15,Donaire161,Barcellona16, Jentschura17}.     It has been predicted that at large interatomic separations, the magnitude of the interaction potential  exhibits spatial  oscillations  \cite{McLone65, Gomberoff66}. In later works, it has been claimed that the potential monotonically decays as  a function of interatomic separation  \cite{Power93,Power95}.   However, the latter result seemed to be in contradiction with the long-distance interaction potential between an excited atom and metal  or dielectric plate, which has been shown theoretically \cite{WylieSipe} and experimentally \cite{Wilson03, Bushev04} to oscillate with the atom-plate distance. 

The reason for the controversy  is divergent energy denominators appearing in time-independent perturbation theory, which can be integrated  by adding an infinitesimal imaginary part  to the divergent denominators,  with its sign determining whether the interaction potential oscillates with the distance or monotonic. In conventional perturbation theory, there is no indication on the correct sign. However, using a dynamic theory with subsequent observation-time averaging,  a third result for the interaction potential on the excited atom, which at long-distance limit oscillates both in magnitude and sign, has been obtained for non-identical atoms \cite{Berman15, Donaireetal15} and later generalised to the case of  identical atoms using a quantum-electrodynamical approach  \cite{Jentschura17}. 

Resolutions of the  contradiction has been offered in a number of recent  publications suggesting that both monotonic and oscillating behaviours are valid, but describe different physical situations involving reversible and irreversible excitation exchange \cite{Milonni15}, or can appear in the same system, where the ground-state atom experiences the monotonic dispersion force, and the excited atom is a subject to the oscillating force \cite{Donaire161,Barcellona16}.    The latter  implies the violation of the action-reaction theorem for the two atoms, but can be justified by taking into account photon emission by the excited atom \cite{Donaire162}.

Another potentially  controversial non-equilibrium situation  can occur in the system of two ground-state atoms out of equilibrium with isotropic EM field, where the monotonic behaviour of the vdW force at large distances has been predicted \cite{Sherkunov09,Sherkunov09C,Behunin10,HAAKH12}, which seems to be in contrast with the oscillating  force on  a ground-state atom out of equilibrium with EM field-dielectric plate system  \cite{Buhmann08,Sherkunov09}.

In this paper, we study the vdW interaction  between two dissimilar atoms  prepared in arbitrary initial states (ground, or excited) out of equilibrium with surrounding isotropic EM field and derive  closed-form expressions for the energy shifts of each atom and  related vdW forces using the Keldysh diagrammatic technique  \cite{Keldysh,LandauX} formulated for few-body  systems \cite{Sherkunov05,Sherkunov07,Sherkunov10}. This method prescribes  regularisation rules of divergent energy denominators  allowing us to avoid controversies associated with the standard time-independent perturbation theory \cite{Berman15}. We  assume that the observation time is smaller than the life-times of the states of the atoms, implying that the atoms experience only virtual transitions,  allowing us to apply a  quasi-stationary version of the theory.  

\begin{figure}[h]
\includegraphics[width=0.4\textwidth]{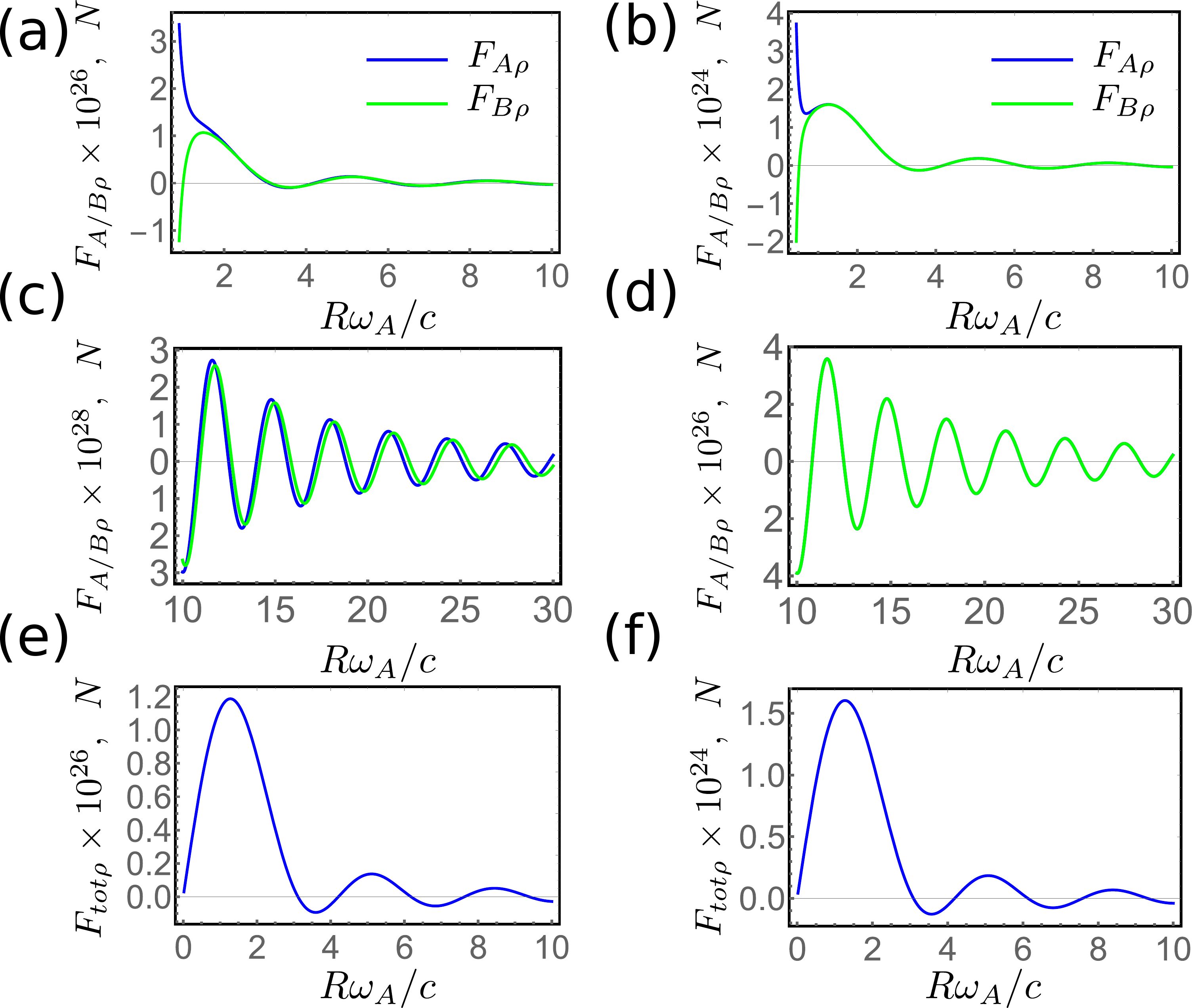}
\caption{ Spatial oscillations of the  vdW forces in the system of two ground-state atoms  out of equilibrium with thermal EM field at $T=\omega_A$.
(a) The vdW forces on atoms $A$, $\mathbf F_A$, and $B$, $\mathbf F_B$, projected  onto the direction $\bm \rho=(\mathbf R_A-\mathbf R_B)/R$ as functions of interatomic separation $R$  for the set of parameters corresponding to optical transitions $5^2S_{1/2}\rightarrow 5^2P_{3/2}$ for  $^{87}Rb$ atom (atom $A$) and   $2^2S_{1/2}\rightarrow 2^2P_{3/2}$ for  $^{40}K$ atom (atom $B$) with optical transition frequencies $\omega_A=1.59eV$ and $\omega_B=1.61eV$.  (b) Same as (a), but for magnetically tuned transition frequencies $(\omega_A-\omega_B)/\omega_A=10^{-4}$. (c)-(d)  Long-distance behaviour of $\mathbf F_{A\rho}$ and $\mathbf F_{B\rho}$ for the set of parameters of (a) and (b). (e)-(f) Net force on the two-atom system $\mathbf F_{tot}=(\mathbf F_A+\mathbf F_B)/2$ projected onto the direction $\bm \rho$ for the set of parameters of (a) and (b). }
\label{Fig1}\end{figure}

We found that in the long-distance regime, $R\gg\lambda$, where $R$ is the interatomic distance and $\lambda$ is a characteristic wavelength of atomic transitions,  both  atoms, in general, experience oscillating and monotonic components of the interaction potentials arising  from resonance emission or absorption of virtual photons by one of the atoms, inducing spatial oscillations for its own potential and the monotonic component for the other atom.  This implies unequal vdW potentials on each atom giving rise to non-reciprocity, which can be explained  when a photon emitted by an excited atom or absorbed by a ground-state atom is taken into account restoring overall momentum balance similarly to what has been shown for  a system of an excited atom and a ground-state one in vacuum \cite{Donaire162}. In the latter case, as we show,  the retarded vdW potential of the ground-state atom looses its oscillating component, while the vdW potential of the excited atom is purely  oscillating in agreement with recent works \cite{Donaire161,Barcellona16}. 
 
Then, we focus on the case  of two ground-state atoms with close transition frequencies, $\omega_A$ and $\omega_B$, surrounded by thermal  EM field, whose photon density does not change much within $\omega_A-\omega_B$. At small interatomic separation $R\ll\lambda$, the interaction is  reciprocal, repulsive,  non-resonant, and the vdW forces decay as $R^{-7}$, as shown in  Fig. \ref{Fig1} (a)-(b). However, at large separations, $R\gg\lambda$,  the system of two atoms becomes non-reciprocal and the vdW forces on each atom are co-directional, almost equal, and resonant. They decay as $R^{-2}$ and oscillate with $R$ almost in-phase (see  Fig. \ref{Fig1} (c)-(d)) giving rise to the sizeable oscillating net force and negligible interatomic force. The former reaches its maximum in the intermediate regime $R\sim\lambda$ (see Fig. \ref{Fig1} (e)-(f)), when the forces on each atom become co-directional and almost equal.

As an example, we numerically calculate the vdW forces in the system of $^{87}Rb$ and $^{40}K$  ground-state atoms out of equilibrium with thermal EM field at temperature close to the dominant transition frequencies of the atoms with and without magnetic field inducing Zeeman splitting. We find that the magnitude of the net force on the atomic system can be within  experimentally available values.

The  vdW forces discussed in this paper can also be induced using artificially created fluctuating light fields  \cite{Brugger15} . We found that the vdW forces not only can be dramatically enhanced, but also controlled and manipulated by applying light fields with tailored spectral properties. As we show in Fig. \ref{Fig2}, in the short-distance regime, the interaction becomes resonantly enhanced provided the energy densities of external EM field, $U(\omega_A)$ and $U(\omega_B)$,  at frequencies $\omega_A$ and $\omega_B$ are not equal. Adjusting the ratio $U(\omega_A)/U(\omega_B)$ would allow one not only to control the magnitudes of the vdW forces, which scale linearly with $U(\omega_A)$ and $U(\omega_B)$, but also change their direction, switching the interaction from repulsive to attractive. In the large-distance regime,  adjusting the spectral densities $U(\omega_A)$ and $U(\omega_B)$ would allow to control the amplitudes of the oscillating components of the vdW forces on each atom,  and even make the interaction monotonic, as shown in Fig. \ref{Fig3}. It would also allow to control the net force on the system, provided the transition frequencies $\omega_A$ and $\omega_B$ are not too close.

 \section{Model}

We consider two dissimilar two-level atoms, $A$ and $B$,  characterised by resonance transition frequencies $\omega_A$ and $\omega_B$ and line-widths $\gamma_A$ and $\gamma_B$, such that $|\omega_A-\omega_B|\gg \gamma_A,\gamma_B$,  located at  positions $\mathbf R_A$ and $\mathbf R_B$ and interacting with isotropic and unpolarised EM field  modelled, in dipole approximation,   by the  Hamiltonian ($\hbar=c=1$), 
\begin{equation}
H_{int}=-\sum_{j=A,B}\int \psi_j^{\dagger}(\mathbf r)\mathbf{d}\cdot \mathbf{E}(\mathbf r)\psi_j(\mathbf r)d^3\mathbf{r}.
\end{equation} 
Here,  $\psi_j(\mathbf r)=\sum_{i=g,e}\phi_i^j(\mathbf r-\mathbf R_j)b_{ji}$ is the field operator of atom $j$,  
\begin{equation}
\mathbf E(\mathbf r)=i\sum_{\mathbf k\mu} \sqrt{\frac{2\pi |\mathbf k|}{V}}\mathbf e_{\mathbf k\mu}\left(\alpha_{\mathbf k\mu}e^{i\mathbf k \mathbf r}-\alpha^{\dagger}_{\mathbf k\mu}e^{-i\mathbf k \mathbf r}\right)
\end{equation}
 is the electric field operator, and  $\mathbf d=e\mathbf r$ is the operator of dipole moment, where $b_{ji}$ is the annihilation operator of the ground ($i=g$) or excited ($i=e$) state of atom $j$ described by the wave-function   $\phi_i^j(\mathbf r-\mathbf R_j)$,  $\alpha_{\mathbf{k}\mu}$ is the annihilation operator of a photon with momentum $\mathbf k$ and polarisation index $\mu$, $e_{\mathbf k\mu}$ is the unit polarisation vector, and $V$ is the quantisation volume.

 At the initial time $t_0$, the atoms are prepared in their initial states $i$ with probabilities $p_i^A$ and $p_i^B$ and are  out of equilibrium with EM field.   We assume that within the observation time  $1/|\omega_A-\omega_B|\ll t_f<\gamma_A^{-1},\gamma_B^{-1}$, the  atoms stay in their initial states and do not equilibrate with the EM field, allowing us to match the vdW potentials with the energy shifts of the initial states of each atom \cite{Donaire161}, and calculate them from the density matrices of atom $j$, 
\begin{eqnarray}
\rho_j(\mathbf r,t;\mathbf r',t')=\Tr[\hat{\psi}_{j}^{\dagger}(\mathbf r',t')\hat{\psi}_j(\mathbf r,t)], \label{rho}
\end{eqnarray} 
where $\hat{\psi}$ is in the Heisenberg picture, using the  adiabatic hypothesis \cite{LandauX}, with the help of Keldysh Green's function method \cite{LandauX,Sherkunov05, Sherkunov07, Sherkunov10}.
 
 \section{van der Waals potential of an atom in a generic surrounding}
 
First we consider a  more general situation when atom $A$ is prepared in an arbitrary state and surrounded by EM field and/or arbitrary magneto-dielectric bodies and calculate its energy shifts. As it was shown in Ref.  \cite{Sherkunov05, Sherkunov07}, the density matrix of atom $A$  is given by the following equation, provided the atom does not change its initial state, \textit{i.e.} the condition $|\omega_A-\omega_B|\gg \gamma_A,\gamma_B$ is fulfilled  (see Appendix A):
\begin{eqnarray}
\rho_A(X,X')&=& \rho_A^0(X,X')e^{-i\langle M_{11}^A\rangle(t-t_0)}e^{i\langle M_{22}^A\rangle(t'-t_0)},\label{rho2}
\end{eqnarray}
where we use $\{X=\mathbf r,t\}$ .
Here 
\begin{eqnarray}
\rho_A^0(X,X')=\phi_i^A(\mathbf r-\mathbf R_A)\phi_i^{A*}(\mathbf r'-\mathbf R_A)e^{-i\epsilon^A_i(t-t')}\label{rho0}
\end{eqnarray}
is the density matrix of non-interacting atom $A$ in state $i$ with bare energy $\epsilon^A_i$  and $M_{11}^A=M_{22}^{A*}$ is the self-energy of atom $A$, 
 \begin{eqnarray}
&& \langle M_{11}^A\rangle=\int d^3\mathbf rd^3\mathbf r'\phi_i^A(\mathbf r-\mathbf R_A)\nonumber\\
 &&\times M_{11}^A(\mathbf r,\mathbf r',\epsilon_i^A)\phi_i^{A*}(\mathbf r'-\mathbf R_A),\label{Mllav}
\end{eqnarray}
with
\begin{eqnarray}
M_{11}^A=i\sum_{\nu \nu '}g_{A11}^{0}(X,X')d^{\nu}d^{\nu '} D_{11}^{\nu \nu '}(X',X),\label{Mll}
\end{eqnarray}
expressed in terms of the atomic propagator, $g_{A11}^0$ and causal  photonic Green's tensor $D_{11}^{\nu\nu'}$. 
The former is defined in terms of vacuum averages  $g_{A11}^{0}(X,X')=-i\langle T\psi_{j}(X)\psi_{j'}^{\dagger}(X')\rangle_{vac}$, and, in the energy domain, takes the form  
\begin{eqnarray}
g_{A11}^0(\mathbf r,\mathbf r',E)=\sum_{i=g,e}\frac{\phi_i^A(\mathbf r-\mathbf R_A)\phi_i^{A*}(\mathbf r-\mathbf R_A)}{(E-\epsilon_i^A+i0)}.
\end{eqnarray}
The latter is defined as
\begin{eqnarray}
D_{11}^{\nu \nu'}(X,X')=-i\langle \hat{T} \hat{E}^{\nu}(X)\hat{E}^{\nu'}(X')\rangle,\label{photonD}
\end{eqnarray}
where $\hat{T}$ is  the time-ordering operator, and $\nu=\{x,y,z\}$.

As it follows from Eq. (\ref{rho2}), the energy shift of atom $A$ induced by the EM field and, thus, the corresponding vdW potential is determined by the real part of the self-energy, $U_A=\Delta\epsilon_i^A=\Re \langle M_{11}^A\rangle$, while the corresponding line-width is equal to its imaginary part $\gamma_i^A=\Im \langle M_{11}^A\rangle$.

Using Eqs. (\ref{rho0}),  (\ref{rho2}), and (\ref{Mll}), we find the vdW potential experienced by atom $A$, 
\begin{eqnarray}
U_A=\Re\left[\frac{i}{2\pi}\int_{-\infty}^{\infty}\frac{d_A^{\nu}d_A^{\nu '}D_{11}^{\nu ' \nu}(\omega,\mathbf R_A,\mathbf R_A)}{\omega-\epsilon_i^A+\epsilon_{\bar{i}}^A}d\omega \right],\label{Uint}
\end{eqnarray}
where $\bar{i}$ stands for the  state of atom $A$, opposite to $i$ and  $d_{A}$ is the transition matrix element of  dipole moment. We use a property of   the photon Green's tensor, $D_{11}^{\nu\nu '}(\mathbf r,\mathbf r',-\omega)= D_{11}^{\nu '\nu }(\mathbf r',\mathbf r,\omega)$, which follows from its definition  (\ref{photonD}) \cite{Sherkunov07} and rewrite (\ref{Uint}) as    $U_A=-\Re[\frac{i}{2\pi}\int_0^{\infty}\tilde{\alpha}_{Ai}^{\nu\nu'}(\omega)D_{11}^{\nu ' \nu}(\omega,\mathbf R_A,\mathbf R_A)d\omega ]$, where $\tilde{\alpha}_{Ag/e}(\omega)=d_{A}^{\nu}d_{A}^{\nu'}\left(\frac{1}{\pm\omega_{A}-\omega-i0}+\frac{1}{\pm\omega_{A}+\omega-i0}\right)$ is related to the polarisability  of atom $A$, 
\begin{eqnarray}
\alpha_{Ag/e}(\omega)=\left(\frac{d_{A}^{\nu}d_{A}^{\nu'}}{\pm\omega_{A}-\omega-i0}+\frac{d_{A}^{\nu}d_{A}^{\nu'}}{\pm\omega_{A}+\omega+i0}\right), \label{polarisability}
\end{eqnarray}
as $\tilde{\alpha}_{Ag/e}(\omega)=\alpha_{Ag/e}(\omega)+2\pi i d_{A}^{\nu}d_{A}^{\nu '}\delta (\omega\pm\omega_{A})$. Note that, for $\omega>0$, the difference between $\tilde{\alpha}$ and $\alpha$ is significant only if atom $A$ is in its excited state, leading to
\begin{eqnarray}
U_A&=&-\Re[\frac{i}{2\pi}\int_0^{\infty}\alpha_{Ai}^{\nu\nu'}(\omega)D_{11}^{\nu ' \nu}(\omega,\mathbf R_A,\mathbf R_A)d\omega ]\nonumber\\
&+&\Re[d_{A}^{\nu}d_{A}^{\nu '}D_{11}^{\nu ' \nu}(\omega_A,\mathbf R_A,\mathbf R_A)p_e^A].\label{def}
\end{eqnarray}
Eq. (\ref{def}) represents the general formula describing the  interaction of a two-level atom prepared in an arbitrary state, excited or ground, with EM field described by the  Green's function $D_{11}$,  provided that the observation time is small compared with the life-time of the atom's initial state. 

To check the result, we calculate the Casimir-Polder force experienced by atom $A$ prepared in an arbitrary state, ground or excited, positioned near a dispersive and absorbing medium. We suppose that the medium is kept at temperature $T$ and is at thermal equilibrium with electromagnetic field. For the initial stage of atom-field interaction, $t<\gamma_A^{-1}$, the atom  does not change its initial state, and the interaction potential experienced by the atom can be evaluated with the help of Eq. (\ref{def}) with $D_{11}=D_r-i\rho_{ph}$ (see Appendix B) and  the photonic density matrix , $\rho_{ph}$, given by the fluctuation-dissipation theorem \cite{LandauIX}: 
\begin{eqnarray}
\rho_{ph}^{\nu \nu'}(\omega,\mathbf r,\mathbf r')=-2N(\omega)\Im D_r^{\nu\nu '}(\omega,\mathbf r,\mathbf r'),\label{rhoph}
 \end{eqnarray}
 where $N(\omega)=\left(e^{\omega/T}-1\right)^{-1}$ is the average  number of photons with frequency $\omega$,  yielding the result of Ref \cite{Buhmann08}:
\begin{eqnarray}
U_i^A&=& U_{eq}+U_{neq},\nonumber\\
U_{eq}&=&-\Re[\frac{i}{2\pi}\int_0^{\infty}(2N(\omega)+1)\alpha_{Ai}^{\nu\nu'}(\omega)\nonumber\\
&\times&D_{r}^{\nu ' \nu}(\omega,\mathbf R_A,\mathbf R_A)d\omega ],\nonumber\\
U_{neq}&=&-\Re[d_{Aji}^{\nu}d_{Aij}^{\nu '}D_{r}^{\nu ' \nu}(\omega_A,\mathbf R_A,\mathbf R_A)\nonumber\\
&\times&[N(\omega_A)p_g^A-(N(\omega_A)+1)p_e^A],\label{cp}
\end{eqnarray}
where the equilibrium potential describing Casimir-Polder interaction of a thermalised atom can be evaluated as $U_{eq} = T\sum_{m=0}^{\infty}(1-\frac{1}{2}\delta_{m0})\alpha_{A}^{\nu'\nu}(i\xi_{m})D_r^{\nu\nu'}(i\xi_{m})$, with  $\xi_m=2\pi mT$ the Matsubara frequency.  Here, we used the property of the polarisability, $\Re[\frac{i}{2\pi}\alpha_A(\omega)]=\delta(\omega-\omega_A)(p_e^A-p_g^A)/2$, which follows from its definition (\ref{polarisability}).

\section{van der Waals interaction between two atoms surrounded by isotopic EM field}
\subsection{General case}
Now we consider the interaction between two atoms, $A$ and $B$, prepared in arbitrary states and embedded in isotopic EM field.
We assume that the optical Stark shift induced by free EM field, as well as Lamb shift due to free EM vacuum fluctuations  are taken into account in the atomic transition frequencies and suppose, without loss of generality, averaging over all possible directions of dipole matrix elements, so that $d_{A/B}^{\nu}d_{A/B}^{\nu'}=\delta_{\nu \nu'}|d_{A/B}|^2/3$, where $\delta$ is the Kronecker symbol. The interaction potential on the atoms is given by Eq. (\ref{def}) with the scattering part of the photon Green's function $D_{11}$, satisfying the equation (see Appendix B):
\begin{widetext}
\begin{eqnarray}
&&D_{11}^{\nu \nu'}(\omega,\mathbf R_A,\mathbf R_A)=-(2N(\omega)+1)\alpha_B^{\nu_1\nu_2}(\omega)D_r^{0\nu \nu_1}(\omega,\mathbf R_A,\mathbf R_B)D_r^{0\nu_2\nu'}(\omega,\mathbf R_B,\mathbf R_A)\nonumber\\
&&+2N(\omega)\Re[\alpha_B^{\nu_1\nu_2}(\omega)D_r^{0\nu \nu_1}(\omega,\mathbf R_A,\mathbf R_B)D_r^{0\nu_2\nu'}(\omega,\mathbf R_B,\mathbf R_A)]+2iN(\omega)\Im[\alpha_B^{\nu_1\nu_2}(\omega)]D_r^{0\nu \nu_1}(\omega,\mathbf R_A,\mathbf R_B)(D_r^{0\nu_2\nu'}(\omega,\mathbf R_B,\mathbf R_A))^*\nonumber\\
&&-2\Im[\alpha_{B}^{\nu_1\nu_2}(\omega)]p_e^BD_r^{0\nu \nu_1}(\omega,\mathbf R_A,\mathbf R_B)(D_r^{0\nu_1\nu'}(\omega,\mathbf R_B,\mathbf R_A))^*.\label{GreensD11}
\end{eqnarray} 
\end{widetext} 
Under these assumptions, with the help of  Eqs. (\ref{GreensD11}) and (\ref{def}),  we find  that apart from  usual equilibrium potential rapidly decaying with interatomic separation\cite{Lifshitz, Lifshitz61,MilonniSmith96} 
\begin{eqnarray}
U_A^{eq}&=&\Re[\frac{i}{\pi}\int_0^{\infty}d\omega (N(\omega)+1/2)\alpha_A(\omega)\alpha_B(\omega)\nonumber\\
&&(D_r^0(\omega,\mathbf R_A-\mathbf R_B))^2\label{Eq}
\end{eqnarray}
describing the interaction between atoms thermalised with EM field, atom $A$ experiences the non-equilibrium resonant potential,
\begin{eqnarray}
&&U_{A}^{neq}=\frac{2|d_A|^2|d_B|^2}{9(\omega_A^2-\omega_B^2)}\left\{\omega_A[N(\omega_B)p_g^B-(N(\omega_B)+1)p_e^B]\right.\nonumber\\
&& \times |D_r^0(\omega_B,\mathbf R_A-\mathbf R_B)|^2-\omega_B[N(\omega_A)p_g^A-(N(\omega_A)+1)p_e^A]\nonumber\\
&&\left.\times\Re[ (D_r^0(\omega_A,\mathbf R_A-\mathbf R_B))^2]\right\}.\label{Uneq}
\end{eqnarray}
disappearing with the equilibration between the atoms and the EM field.  Indeed, assuming that the EM field is thermal, \textit{i.e} obeys Bose-Einstein distribution, and the probabilities to find each atom in a specific state are described by Boltzmann distribution, $p_g^j=e^{\omega_j/T}\left(e^{\omega_j/T}+1\right)^{-1}$ and $p_e^j=p_g^j  e^{-\omega_j/T}$, the non-equilibrium potential vanishes.

Using the same procedure, we find, that the non-equilibrium vdW potential for atom $B$,
\begin{eqnarray}
&&U_{B}^{neq}=-\frac{2|d_A|^2|d_B|^2}{9(\omega_A^2-\omega_B^2)}\left\{\omega_B[N(\omega_A)p_g^A-(N(\omega_A)+1)p_e^A]\right.\nonumber\\
&& \times |D_r^0(\omega_A,\mathbf R_A-\mathbf R_B)|^2-\omega_A[N(\omega_B)p_g^B-(N(\omega_B)+1)p_e^B]\nonumber\\
&&\left.\times \Re[(D_r^0(\omega_B,\mathbf R_A-\mathbf R_B))^2]\right\}.\label{UneqB}
\end{eqnarray}
is not, in general, equal to $U_A^{neq}$. 
Moreover, in the long-distance regime, $R\gg\lambda$, they both contain oscillating and monotonic in $R$ terms, which can be seen by substituting $\Re[(D_r^0)^2]$ and $|D_r^0|^2$ \cite{LandauIX},
\begin{eqnarray}
&&|D_r^0(\omega,R)|^2=\frac{2\omega^4}{R^2}\left(1+1/(\omega R)^2+3/(\omega R)^4\right),\label{Dm}\\
&&\Re(D_r^0(\omega,R))^2=\frac{2\omega^4}{R^2}\left[\cos(2\omega R)\left(1\right.\right.\nonumber\\
&&\left.-5/(\omega R)^2+3/(\omega R)^4)\right)\nonumber\\
&&\left.+\sin(2\omega R)\left(3/(\omega R)^3-1/(\omega R)\right)\right],\label{D}
\end{eqnarray}
into (\ref{Uneq}) and  (\ref{UneqB}),
\begin{widetext}
\begin{eqnarray}
&&U_{A}^{neq}=\frac{4|d_A|^2|d_B|^2\omega_A\omega_B}{9R^2(\omega_A^2-\omega_B^2)}\left\{\omega_B^3[N(\omega_B)p_g^B-(N(\omega_B)+1)p_e^B] -\omega_A^3[N(\omega_A)p_g^A-(N(\omega_A)+1)p_e^A]\cos(2\omega_AR) \right\},\label{UneqL}\\
&&U_{B}^{neq}=-\frac{4|d_A|^2|d_B|^2\omega_A\omega_B}{9R^2(\omega_A^2-\omega_B^2)}\left\{\omega_A^3[N(\omega_A)p_g^A-(N(\omega_A)+1)p_e^A] -\omega_B^3[N(\omega_B)p_g^B-(N(\omega_B)+1)p_e^B] \cos(2\omega_BR)\right\}.\label{UneqBL}
\end{eqnarray}
\end{widetext}
The origin of these components depends  on which atom takes part in resonance processes: for the potential on atom $A$, the oscillations are due to \textit{its} spontaneous (stimulated) emission of virtual quanta or \textit{its} resonant absorption of an external photon, while the monotonic component is due to the resonance processes involving    \textit{atom $B$} and vice versa. However, in the short-distance regime, $R\ll \lambda$, the oscillations disappear and Eqs. (\ref{Uneq}) and (\ref{UneqB})  give us monotonic and  equal vdW potentials:
\begin{widetext}
\begin{eqnarray}
U_{A/B}^{neq}=\frac{6|d_A|^2|d_B|^2}{9R^6(\omega_A^2-\omega_B^2)}[N(\omega_B)p_g^B-(N(\omega_B)+1)p_e^B -N(\omega_A)p_g^A+(N(\omega_A)+1)p_e^A] .\label{UAS}
\end{eqnarray}
\end{widetext}

 If one of atom $A$ is excited and atom $B$ is in its ground state and the external EM field is absent, the long-distance vdW potential of the excited atom Eqs. (\ref{UneqL}) exhibits spatial  oscillations  both in sign and magnitude supporting the results of Ref. \cite{Donaireetal15}  and the one of the ground-state atom (\ref{UneqBL}) is monotonic  in agreement with  Refs \cite{Donaire161,Barcellona16}. In this case, the asymmetry leading to non-reciprocal vdW forces violating the action-reaction theorem has been attributed to a net transfer of linear momentum to the quantum fluctuations of the EM field due to spontaneous emission by the excited atom \cite{Donaire162}.

\subsection{Two ground-state atoms in isotopic EM field} 
Next, we consider two ground-state atoms out of equilibrium with  external EM field. For the short-distance regime $R\ll\lambda$ the non-equilibrium vdW potentials can be found from Eqs. (\ref{UAS}):
\begin{eqnarray}
U_{A/b}^{neq}=\frac{4|d_A|^2|d_B|^2[\omega_AN(\omega_B)-\omega_BN(\omega_A)]}{3R^6(\omega_A^2-\omega_B^2)},\label{nonret}
\end{eqnarray}
and  can be related to the field assisted vdW forces acting along the direction $\bm \rho=(\mathbf R_A-\mathbf R_B)/R$, $\mathbf F_{A}=-\nabla_{A}U_A^{neq}$ and $\mathbf F_{B}=-\nabla_{B}U_B^{neq}$, 
 \begin{eqnarray}
&&\mathbf F_A=-\mathbf F_B\nonumber\\
&=&\frac{8|d_A|^2|d_B|^2[\omega_AN(\omega_B)-\omega_BN(\omega_A)]\bm \rho}{R^7(\omega_A^2-\omega_B^2)}.\label{nonret}
\end{eqnarray}
In the large distance regime, $R\gg\lambda$, we find 
\begin{eqnarray}
U_A^{neq}&=&\frac{4|d_A|^2|d_B|^2\omega_A\omega_B}{9R^2(\omega_A^2-\omega_B^2)}\nonumber\\
&\times&[\omega_B^3N(\omega_B)-\omega_A^3N(\omega_A)\cos(2\omega_AR)],\label{retA}\\
U_B^{neq}&=&-\frac{4|d_A|^2|d_B|^2\omega_A\omega_B}{9R^2(\omega_A^2-\omega_B^2)}\nonumber\\
&\times&[\omega_A^3N(\omega_A)-\omega_B^3N(\omega_B)\cos(2\omega_BR)],\label{retB}
\end{eqnarray}
which leads to:
\begin{eqnarray}
\mathbf F_A&=-&\frac{8|d_A|^2|d_B|^2N(\omega_A)\omega_A^5\omega_B\bm{\rho}}{9R^2(\omega_A^2-\omega_B^2)}\sin(2\omega_AR),\label{retAF}\\
\mathbf F_B&=-&\frac{8|d_A|^2|d_B|^2N(\omega_B)\omega_A\omega_B^5\bm{\rho}}{9R^2(\omega_A^2-\omega_B^2)}\sin(2\omega_BR).\label{retBF}
\end{eqnarray}

\subsection{Two ground-state atoms in thermal EM field} 
In the case of thermal EM field at temperature $T\gg |\omega_A-\omega_B|$, so that $N=N(\omega_A)\approx N(\omega_B)$, the short-distance forces described by Eq. (\ref{nonret}) are repulsive, have equal magnitudes and non-resonant (see Fig. \ref{Fig1} (a)-(b)). Consequently, the net force is absent. However, in the large-distance case, the forces given by Eqs. (\ref{retAF}) and (\ref{retBF}) are resonant,  have the same direction and amplitude, and  show spatial oscillations almost in-phase (see  Fig. \ref{Fig1} (c)-(d)) giving rise  to  spatially oscillating net force  $\mathbf F_{tot}=(\mathbf F_A+\mathbf F_B)/2$,  as shown in  Fig. \ref{Fig1} (e)-(f).  In the intermediate regime  $R \sim \lambda$, in which the interaction crosses over from mutual  monotonic repulsion to  spatial oscillations, the net force reaches its maximum with its direction   towards the atom with smaller transition frequency, as we show in  Fig. \ref{Fig1} (e)-(f). At the same time, the vdW forces on each atom become almost equal in direction and magnitude.  Note that in the long-distance regime, the equilibrium contribution to the field assisted vdW force, $\mathbf F_A=-\mathbf F_B=-\frac{4T|d_A|^2|d_B|^2}{\omega_A\omega_BR^7}\bm{\rho}$,\cite{LandauIX} can be neglected.

 As an example, we consider a system of  $^{87}Rb$ and $^{40}K$  atoms prepared in $5^2S_{1/2}$ and $2^2S_{1/2}$ grounds states respectively out of equilibrium with thermal EM field at temperatures comparable with the quasi-resonant  transition energies for $5^2S_{1/2}\rightarrow 5^2P_{3/2}$ of the $^{87}Rb$, $\omega_A=1.59 eV$, and $2^2S_{1/2}\rightarrow 2^2P_{3/2}$ of the $^{40}K$ atoms, $\omega_B=1.61eV$ and calculate the net vdW force numerically (Fig. \ref{Fig1} (e)). However, the magnitude of the net force appears to be too small to be detected experimentally. Applying  external magnetic field would result in Zeeman shifts of the atomic energy levels allowing one to tune the transition frequencies and enhance the resonant net force.  For the relative detuning $\delta\omega=|\omega_A-\omega_B|/\omega_A=10^{-4}$ limited by the Doppler broadening  $\Delta\omega\approx 10^{-5}\omega_A$, we found  that the maximum  value of the net force $F_{tot}^{max}\approx  10^{-23}N$ (see Fig. \ref{Fig1} (f)),  which is within experimentally achievable values \cite{Schreppler2014}.
 
 \subsection{Two ground-state atoms in artificial random EM field} 
 \begin{figure}[h]
\includegraphics[width=0.49\textwidth]{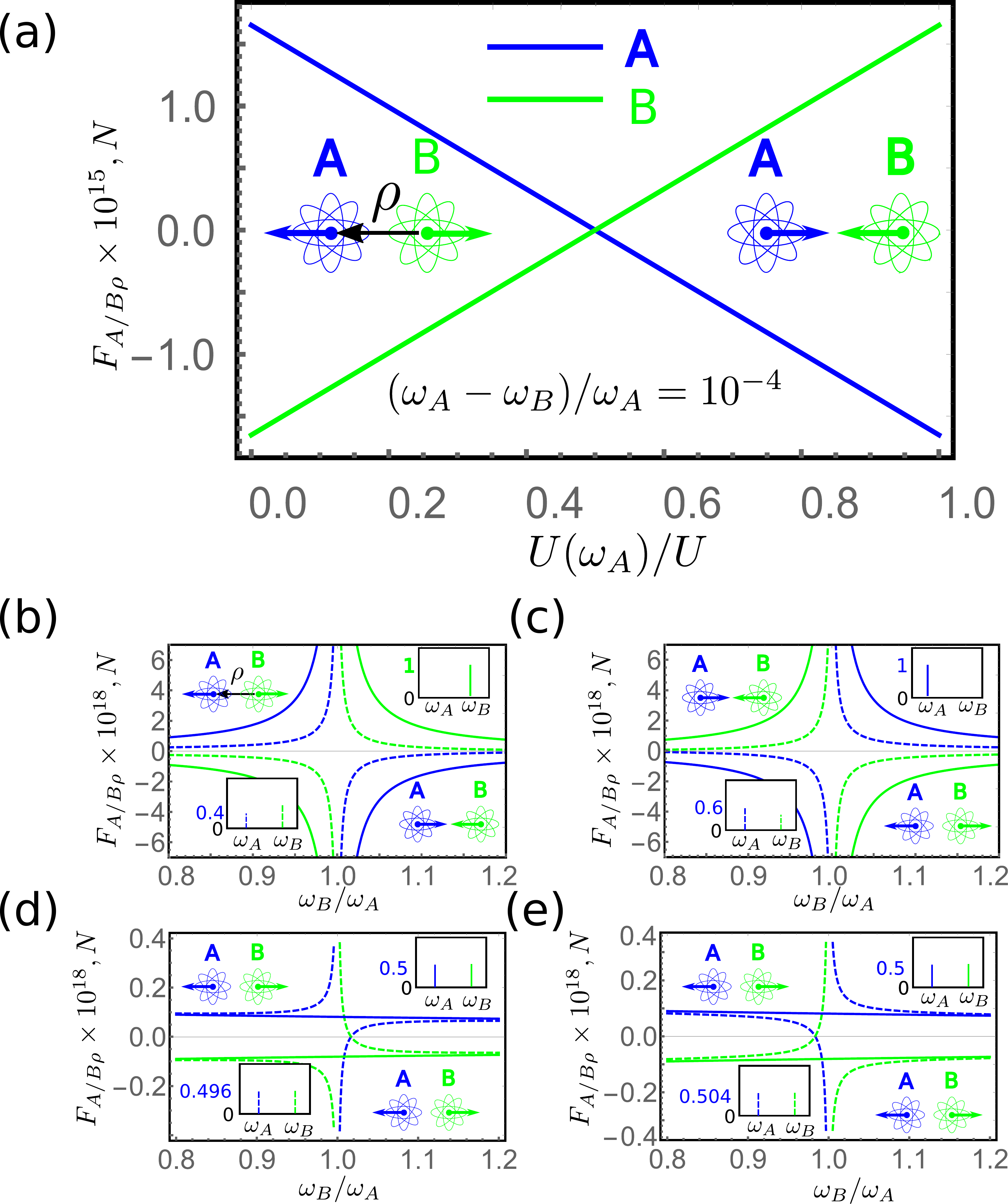}
\caption{ Enhancement of the short-distance van der Waals forces on two ground-state atoms by artificially created  random light  for magnetically tuned  optical transitions $5^2S_{1/2}\rightarrow 5^2P_{3/2}$ of  $^{87}Rb$ atom (atom A) and   $2^2S_{1/2}\rightarrow 2^2P_{3/2}$ of  $^{40}K$ atom (atom B). The atoms separated by a distance $R=0.3\lambda$ are out of equilibrium with artificial random light with narrow spectral energy  densities peaked at $\omega_A$ and $\omega_B$ (see insets).  (a) The van der Waals forces as  functions of photon energy density $U(\omega_A)/U$, where $U=U(\omega_A)+U(\omega_B)$, for the transition frequencies $(\omega_A-\omega_B)/\omega_A=10^{-4}$. (b)-(e) The van der Waals forces as functions of $\omega_B/\omega_A$ for a set of photon energy densities  shown in insets. Insets: Photon energy densities, $U(\omega_A)$ and $U(\omega_B)$, normalised to $U=0.6\times 10^{-3}J/m^3$   }
\label{Fig2}\end{figure} 

\begin{figure}[h]
\includegraphics[width=0.49\textwidth]{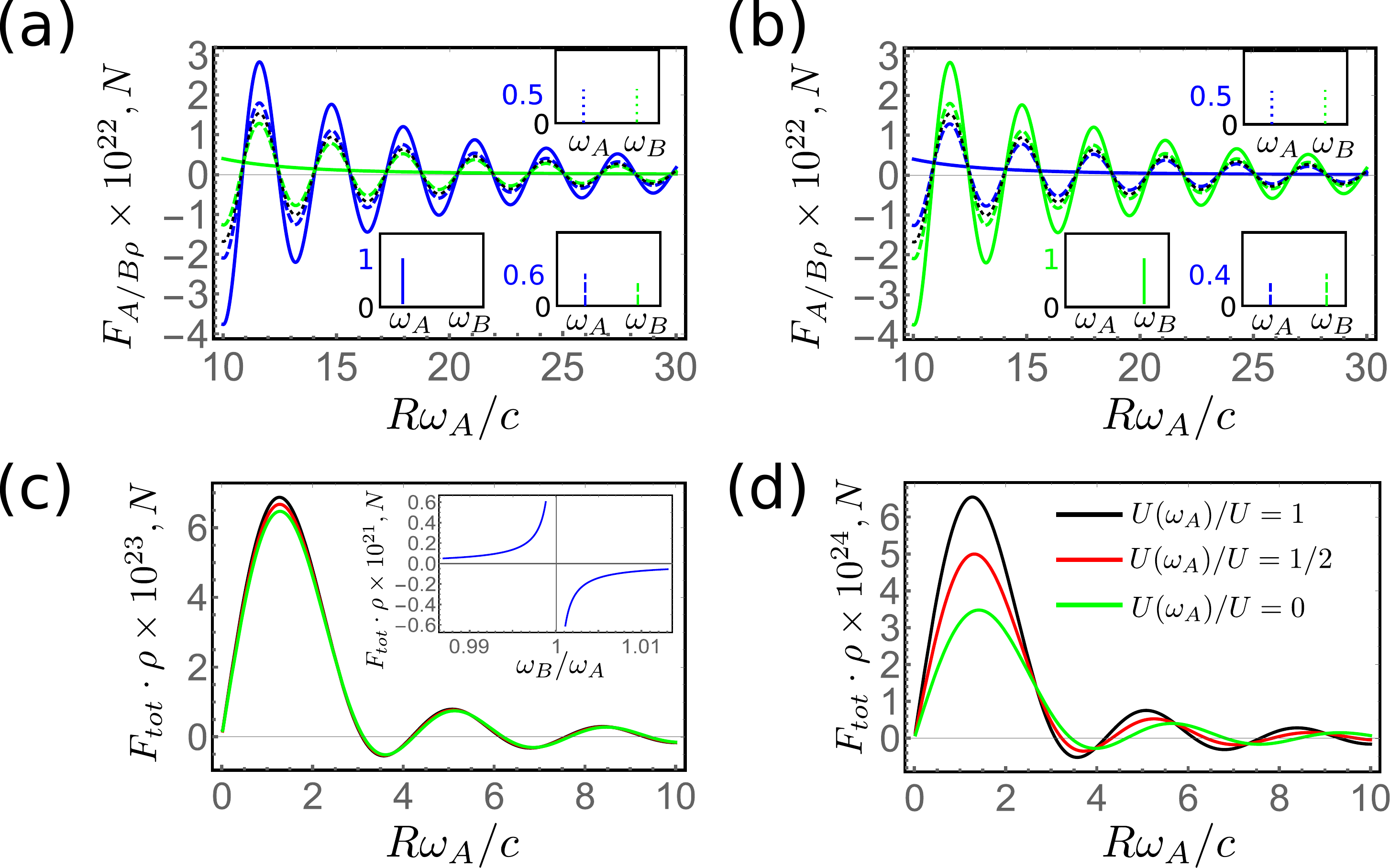}
\caption{(a)-(b) Long-distance van der Waals forces on atoms $A$ and $B$ (see caption of Fig. \ref{Fig2}) as functions of interatomic separation, $R$, for $(\omega_A-\omega_B)/\omega_A=10^{-4}$ and $U(\omega_A)/U$  ($U(\omega_B)/U$) shown in insets. (c) Net force on the system of the two atoms as a function of $R$ for  $(\omega_A-\omega_B)/\omega_A=10^{-2}$ for the set of photon energy densities shown in (d). Inset: Net force as a function of $\omega_B/\omega_A$. (d) Same as (c), but  $(\omega_A-\omega_B)/\omega_A=10^{-1}$.    }
\label{Fig3}\end{figure}

 The forces discussed in this paper can be induced  not only by thermal EM field, but also using  artificially created random isotopic light in a small cavity \cite{Brugger15}. This would allow not only to enhance  the vdW forces compared with the thermal light, but also to control and manipulate their direction and magnitude.  To demonstrate this point, we consider two ground-state  atoms in a small cavity  filled with random light characterised by energy density, $U(\omega)$,  peaked at $\omega_A$ and $\omega_B$ (see insets of Fig. \ref{Fig2}). We calculate the vdW forces on the atoms numerically,  for a cavity of volume $V=(50\mu m)^3$ and the light generated by a laser with the power $P=0.5mW$ corresponding to the total energy density of random light in the cavity, $U=U(\omega_A)+U(\omega_B)\approx6\times 10^{-4}J/m^3$ \cite{Brugger15}. As in the previous example, we choose  a system of  $^{87}Rb$ and $^{40}K$  atoms prepared in $5^2S_{1/2}$ and $2^2S_{1/2}$ respectively.
 
 At small interatomic separations, $R\ll\lambda$, the vdW forces on each atom are equal in magnitude, however their direction depends  on the ratios of $U(\omega_A)/U(\omega_B)$ and  $\omega_A/\omega_B$ as shown in Fig. \ref{Fig2}. For $U(\omega_A)=0$, when all the photonic energy density is concentrated at the frequency $\omega_B$, the  interaction is repulsive provided  $\omega_A>\omega_B$ and attractive for $\omega_A<\omega_B$ (see Fig. \ref{Fig2} (a),(b) and (c)). As the ratio $U(\omega_A)/U(\omega_B)$ increases, the magnitude of the forces decreases linearly taking their minimum at $U(\omega_A)=U(\omega_B)$. Further increase of $U(\omega_A)/U(\omega_B)$ is accompanied by the linear increase of the magnitudes of the forces, however, the interaction becomes attractive for $\omega_A>\omega_B$ and repulsive otherwise.  For $U(\omega_A)\neq U(\omega_B)$, the interaction demonstrates resonance behaviour at $\omega_A\approx\omega_B$, as shown in Fig. \ref{Fig2} (b) and (c), however, as $U(\omega_A)$ approaches $U(\omega_B)$, the forces become repulsive independently of the ratio $\omega_A/\omega_B$ and non-resonant, as shown in Fig. \ref{Fig2} (d) and (e), in agreement with Eq. (\ref{nonret}). Note that the amplitudes of the forces induced by artificial random light can be up to nine orders of magnitude greater than the ones induced by thermal light discussed above.
 
At large interatomic separations, $R\gg\lambda$, the vdW forces on each atom are resonant, have the same direction and oscillate with the interatomic separation almost in phase, however, their amplitudes depend on the ratio $U(\omega_A)/U(\omega_B)$, as shown in Fig. \ref{Fig3} (a) and (b).  At $U(\omega_B)=0$,  the force on atom $B$ looses its oscillating component and drops with the interatomic distance as  $R^{-3}$, in agreement with \cite{Sherkunov09}, while the oscillation amplitude of the force on  atom $A$  takes its maximum value (see Fig. \ref{Fig3} (a)). As the ratio $U(\omega_B)/U(\omega_A)$ increases, the oscillation amplitude of atom $A$ decreases, while  it increases for atom $B$ to equalise at $U(\omega_A)=U(\omega_B)$,  in agreement with Eqs. (\ref{retAF}) and (\ref{retBF}).  Further increase in $U(\omega_B)/U(\omega_A)$ leads to the decrease of the oscillation component of atom $A$, which disappears at $U(\omega_A)=0$, as shown in Fig. \ref{Fig3} (b).  Again, as in the case of thermal EM field, the artificial random radiation generates a net specially oscillation force on the system of two atoms, which takes its maximum at $R\sim \lambda$. However, in the vicinity of the resonance $\omega_A=\omega_B$ (see inset of Fig. \ref{Fig3} (c)),  the net force is determined by the total energy density $U=U(\omega_A)+U(\omega_B)$, but not by $U(\omega_A)$ and $U(\omega_B)$, as shown in Fig. \ref{Fig3} (c). Thus, in order to control the net force, one has to detune from the resonance, as we demonstrate in Figs. \ref{Fig3}  (d).

\section{Conclusions}

Finally, we comment on the disagreement with previously found monotonic long-distance vdW potential between atoms out of equilibrium with EM field \cite{Sherkunov09,Sherkunov09C,Behunin10,HAAKH12} where, the interaction potential was \textit{a priori} assumed equal for each atom and interpolated from the calculations for the atom with vanishing absorption rate. However, as we show in this work, this procedure is not sufficient if the absorption rates of both atoms are not small.

In this paper, we presented  new formula for the vdW potential in the system of an atom surrounded by arbitrary magneto-dielectric bodies and EM field. We applied this formula to the case of two atoms prepared in arbitrary states out of equilibrium with EM field. We found, that in the long-distance regime, the vdW potentials  have both monotonic and oscillating behaviour with interatomic distance and, in general, unequal for each atom resulting in the net resonant spatially oscillating force. We suggest that the vdW forces can be controlled and manipulated with the help of artificially created random light with tailored spectral properties.  In the particular case of a system with an excited atom and a ground-state one in EM vacuum, our results are  in agreement with the recent findings reported in Refs. \cite{Donaire161, Donaire162, Barcellona16}.

\appendix

\numberwithin{equation}{section}
\numberwithin{figure}{section}
\renewcommand{\theequation}{A.\arabic{equation}}
\renewcommand{\thefigure}{A.\arabic{figure}}

\setcounter{equation}{0}
\setcounter{figure}{0}

\section{Appendix A: Derivation of Eq. (\ref{rho2})}

\begin{figure}[h]
\includegraphics[width=0.4\textwidth]{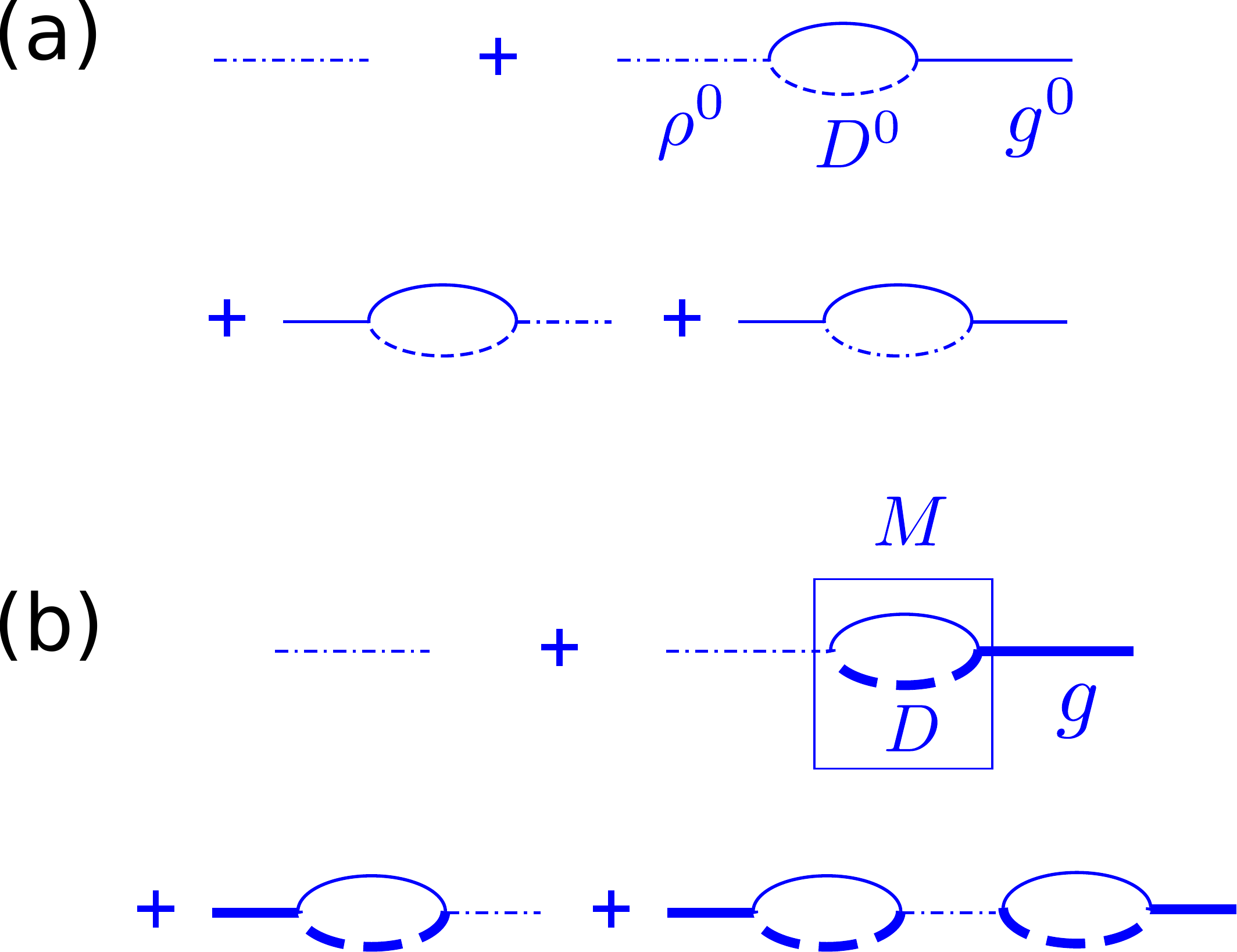}
\caption{(a) Feynman diagrams describing the dressing of the density matrix of atom $j$, $\rho_j^0$ (dashed-dotted line) by electromagnetic field (first three diagrams) and field-induced transition (last diagram) up to the second order perturbation theory. The bare propagator of atom $j$, $g_{jll'}^{0}$ is shown as thin solid line and the bare photon Green's tensor $D_{ll'}^0$ is shown as thin dashed line. 
 The summation over Keldysh indices   $l=1,2$ is assumed in each vertex. (b)  Summation of Feynman diagrams for  the density matrix. Thick lines correspond to the total Green's functions. }
\label{Fig4}\end{figure}

In the interaction picture, the Keldysh Green's functions for atom $j$, 
\begin{eqnarray}
G_{ll'}^j(X,X')=-i\langle T_c \psi_{jl}(X)\psi_{jl'}^{\dagger}(X')S_c(t_f,t_0)\rangle,
\end{eqnarray}
and EM field, 
\begin{eqnarray}
D_{ll'}^{\nu \nu'}(X,X')=-i\langle T_c E_l^{\nu}(X)E_{l'}^{\nu'}(X')S_c(t_f,t_0)\rangle,\label{Ddef}
\end{eqnarray}
 where $X=\{\mathbf r,t\}$ and $\nu=x,y,z$ describes the projection on the corresponding axis,   are defined on the Keldysh contour, which goes in time from $-\infty$ to $\infty$ for $l=1$ and from $\infty$ to $-\infty$ for $l=2$ determining the (anti-) chronological ordering $T_c$ \cite{LandauX,Sherkunov07}. The time-evolution operator, $S_c(t_f,t_0)=T_c \exp[i\sum_{l=1,2}(-1)^l\int_{t_0}^{t_f}dtH_{int,l}(t)]$, can be expanded in $H_{int}$ enabling one  to construct the perturbation series  for the density matrix of atom $j$, $\rho_j=iG_{12}^j$.  Applying the exact Wick's theorem to the atomic operators \cite{Sherkunov05, Sherkunov07} $T_c\psi_{jl}(X)\psi_{jl'}^{\dagger}(X')=:\psi_{jl}(X)\psi_{jl'}^{\dagger}(X'):+ig_{j,ll'}^0(X,X')$, where $:...:$ means normal ordering and  the atomic propagator is determined in terms of vacuum average 
$g_{jll'}^{0}(X,X')=-i\langle T_c\psi_{jl}(X)\psi_{jl'}^{\dagger}(X')\rangle_{vac}$, 
we find the perturbation series, as shown in Fig. \ref{Fig4} (a), where the first Feynman diagram describes non-interacting atom $j$, the second and third diagrams correspond to the elastic scattering of EM field on atom $j$, and the fourth diagram describes spontaneous emission or resonant absorption of a photon. Under the condition $t\ll\gamma_j^{-1}$, we can neglect the fourth term. 

Summing up all relevant reducible bubble diagrams giving rise to atom-EM field interactions, we arrive at the density matrix of atom $j$ described by the Feynman diagrams depicted in Fig.\ref{Fig4}  (b)\cite{Sherkunov05,Sherkunov07}:
\begin{eqnarray}
&&\rho_j(X,X')=\rho_j^0(X,X')\nonumber\\
&+&\int dX_1dX_2\rho_j^0(X,X_1)M_{22}^j(X_1,X_2)g_{j22}(X_2,X')\nonumber\\
&+&\int dX_1dX_2g_{j11}M_{11}^j\rho_j^0\nonumber\\
&+&\int dX_1dX_2dX_3dX_4g_{j11}M_{11}^j\rho_j^0M_{22}^jg_{j22},\nonumber\\
&&g_{jll'}=g_{jll'}^{0}+\sum_{l_1,l_2}\int dX_1dX_2g_{jll_1}^{0}M_{l_1l_2}^jg_{jl_2l'},\label{Dysong}
\end{eqnarray}   
where we omit obvious arguments and  $\rho^0$ and $M_{11}^j=M_{11}^{j*}$  are  given by (\ref{rho0}) and (\ref{Mll}) respectively. 
Keeping in mind, that atom $j$ does not change its initial state $i$ during the interaction with EM field, we factorise the density matrix 
\begin{eqnarray}
\rho_j(X,X')=\phi_i^j(\mathbf r-\mathbf R_A)f(t)\phi_i^{j*}(\mathbf r'-\mathbf R_A)f^*(t')\label{fact}
\end{eqnarray}
in terms of the wave functions of non-interacting atom $j$, where $f(t)$ obeys the equation \cite{Sherkunov05,Sherkunov07}:
\begin{eqnarray}
i\frac{\partial f(t)}{\partial t}-\epsilon_i^j f(t)=\int_{t_0}^{\infty}\langle M_{11}^j(t,t')\rangle f(t'),\label{f}
\end{eqnarray}
and $\langle M_{11}^j(t,t')\rangle$ is given by (\ref{Mllav}). Eq. (\ref{f}) can be solved in the pole approximation,
\begin{eqnarray}
f(t)=e^{-i\epsilon_i^j t}e^{-i\langle M_{11}^j(\epsilon_i^j)\rangle (t-t_0)},
\end{eqnarray}
which, along with Eq. (\ref{fact}), yields Eq. (\ref{rho2}).

\section{Appendix B: Derivation of Eq. (\ref{GreensD11})}
\numberwithin{equation}{section}
\renewcommand{\theequation}{B.\arabic{equation}}
In the presence of atom $B$, the photon Green's functions (\ref{Ddef}) can be calculated  in the lowest orders of perturbation theory \cite{Sherkunov05,Sherkunov07}, 
\begin{eqnarray}
&&D_{ll'}^{\nu \nu '}(X,X')=D_{ll'}^{0\nu \nu '}(X,X')+\int dX_1dX_2D_{ll_1}^{0\nu \nu_1}(X,X_1)\nonumber\\
&&\times \Pi_{Bl_1l_2}^{\nu_1\nu_2}(X_1,X_2)D_{0l_2l'}^{\nu_2\nu'}(X_2,X'),
\end{eqnarray}
with the polarisation operators
 \begin{eqnarray}
&&\Pi_{Bll'}^{\nu\nu'}(X_1,X_2)=(-1)^{l+l'}d^{\nu}d^{\nu'}[g_{Bll'}^0(X_1,X_2)\rho_B^0(X_2,X_1)\nonumber\\
&&+\rho_B^0(X_1,X_2)g_{Bl'l}^0(X_2,X_1)],\label{Pidef}
\end{eqnarray}
 where summation over repeating indices is assumed.  However, only three Green's functions are linearly independent, allowing us to express the energy shifts in terms of the retarded, $D_{r}$, and advanced, $D_a=(D_r)^*$,   Green's function, and  the photon density matrix $\rho_{ph}^{\nu \nu'}=iD_{12}^{\nu \nu'}$ satisfying the equations \cite{LandauX,Sherkunov07}:
\begin{equation}
D_{11}=D_r-i\rho_{ph},\label{D111}
\end{equation}
\begin{equation*}
D_r^{\nu \nu'}(\omega,\mathbf R_A,\mathbf R_A)=D_r^{0\nu \nu'}(\omega,\mathbf R_A,\mathbf R_A)\nonumber\\
\end{equation*}
\begin{equation}
+D_r^{0\nu \nu_1}(\omega,\mathbf R_A,\mathbf R_B)\Pi_{Br}^{\nu_1 \nu_2}(\omega)D_r^{0\nu_2\nu'}(\omega,\mathbf R_B,\mathbf R_A),\label{Dr}
\end{equation}
\begin{equation}
\rho_{ph}=\rho_{ph}^0+D_r^0\Pi_r\rho_{ph}^0+\rho_{ph}^0\Pi_aD_a^0-iD_r^0\Pi_{12}D_a^0,\label{rhoo}
\end{equation}
where the free photon density matrix for isotopic and unpolarised EM field with occupation numbers $N(\omega)$  obeys the fluctuation-dissipation relation  $\rho_{ph}^{0\nu \nu'}(\omega,\mathbf r,\mathbf r')=-2N(\omega)\Im D_r^{0\nu\nu '}(\omega,\mathbf r,\mathbf r')$ \cite{LandauIX,Sherkunov09} and the polarisation operators obey the equations
\begin{eqnarray}
\Pi_r=\Pi_{11}+\Pi_{12},\label{Pir}\\
\Pi_a=\Pi_{11}+\Pi_{21},\label{Pia}
\end{eqnarray}
Direct calculations with the help of Eqs. (\ref{polarisability}) and (\ref{Pidef}) reveals: 
\begin{eqnarray}
&&\Pi_{Br}^{\nu_1 \nu_2}(\omega)=-\alpha_B^{\nu_1\nu_2}(\omega),\;\Pi_a^{\nu_1\nu_2}=\Pi_r^{\nu_2\nu_1*},\;\label{Pir1}\\
&&\Pi_{B12}^{\nu_1 \nu_2}(\omega)=-2\Im[\alpha_B^{\nu_1\nu_2}(\omega)]p_e^B.\label{Pi121}
\end{eqnarray}
Thus, plugging Eqs. (\ref{Pir1}) and (\ref{Pi121}) into (\ref{D111}) - (\ref{rhoo})) leads to Eq. (\ref{GreensD11}).

\bibliography{ybs.bib}{}
\bibliographystyle{apsrev}
\end{document}